\documentclass[pra,twocolumn,amsmath,amssymb,superscriptaddress,longbibliography]{revtex4-1}

\usepackage{bm}
\usepackage{graphicx}
\usepackage{braket}
\usepackage{url}
\usepackage{datetime}
\usepackage{stackrel}
\usepackage{booktabs}
\usepackage{array}
\usepackage{booktabs}
\usepackage{xcolor}

\newcommand{\myb}{\textbf}
\newcommand{\bea}{\begin{eqnarray}}
\newcommand{\eea}{\end{eqnarray}}
\newcommand{\dt}{\partial_t}
\newcommand{\dz}{\partial_z}
\newcommand{\dx}{\partial_x}

\newcommand{\Oz}{\text{O}_3}

\newcommand{\bwe}{\begin{widetext}\begin{eqnarray}}
\newcommand{\ewe}{\end{eqnarray}\end{widetext}}
\newcommand{\bi}{\begin{itemize}}
\newcommand{\ei}{\end{itemize}}
\newcommand{\degree}{$^{\circ}\,$}
\newcommand{\degreens}{$^{\circ}$}

\begin{document}

\title{Photochemically-induced acousto-optics in gases}

\author{P. Michel}
\affiliation{Lawrence Livermore National Laboratory, Livermore, CA 94551, USA}
\author{L. Lancia}
\affiliation{LULI -- CNRS, CEA, Sorbonne Universit\'e, Ecole Polytechnique, Institut Polytechnique de Paris, F-91128 Palaiseau, France}
\author{A. Oudin}
\affiliation{Lawrence Livermore National Laboratory, Livermore, CA 94551, USA}
\author{E. Kur}
\affiliation{Lawrence Livermore National Laboratory, Livermore, CA 94551, USA}
\author{C. Riconda}
\affiliation{LULI, Sorbonne Universit\'e, CNRS, Ecole Polytechnique, CEA, F-75252 Paris, France}
\author{K. Ou}
\author{V. M. Perez-Ramirez}
\author{J. Lee}
\author{M. R. Edwards}
\affiliation{Stanford University, Stanford, CA 94305, USA}

\begin{abstract}
Acousto-optics consists of launching acoustic waves in a medium (usually a crystal) in order to modulate its refractive index and create a tunable optical grating. In this article, we present the theoretical basis of a new scheme to generate acousto-optics in a gas, where the acoustic waves are initiated by the localized absorption (and thus gas heating) of spatially-modulated UV light, as was demonstrated in Y. Michine and H. Yoneda, {\it{Commun. Phys.}} {\bf{3}}, 24 (2020). We identify the chemical reactions initiated by the absorption of UV light via the photodissociation of ozone molecules present in the gas, and calculate the resulting temperature increase in the gas as a function of space and time. Solving the Euler fluid equations shows that the modulated, isochoric heating initiates a mixed acoustic/entropy wave in the gas, whose high-amplitude density (and thus refractive index) modulation can be used to manipulate a high-power laser. We calculate that diffraction efficiencies near 100\% can be obtained using only a few millimeters of gas containing a few percent ozone fraction at room temperature, with UV fluences of less than 100 mJ/cm$^2$---consistent with the experimental measurements. Our analysis suggests possible ways to optimize the diffraction efficiency by changing the buffer gas composition. Gases have optics damage thresholds two to three orders of magnitude beyond those of solids; these optical elements should therefore be able to manipulate kJ-class lasers.
\end{abstract}

\maketitle

\section{Introduction}

Acousto-optics is a well-known method used to control and shape lasers pulses \cite{BoydBook}. It consists of launching a sound wave in a medium (usually a crystal), for example using a piezo-electric transducer; the pressure from the wave modulates the refractive index in the crystal, creating a tunable optical grating. The same process can be achieved in a gas. While acousto-optics effects in gases have been known for a long time \cite{DurrIJIMW86}, their potential for high-power lasers and their applications was only recently outlined  \cite{MichineCP20,SchrodelNP23}. The key benefit of using gases is their higher optics damage threshold compared to solids, typically by two to three orders of magnitude; for example, a damage threshold of 1.6 kJ/cm$^2$ was measured in Ref. \cite{MichineCP20}. Since optics damage is what dictates the physical size (and in large part the cost) of high-power laser facilities, gas optics could be transformational for the area of high-power lasers and their applications, including inertial confinement fusion (ICF) or inertial fusion energy (IFE). Compared to plasma optics, which also aim at manipulating high-power lasers beyond the damage threshold of solids \cite{EdwardsPRA22,EdwardsPRL22,LehmannPRL16,MilchbergPT19,MichelPRL14,PalastroPOP15,PengPRA21}, gases offer the advantage of being generally easier to control; like plasma optics, they can also be transient and re-created at high repetition rate. This makes them a potentially very attractive solution to the problem of the final optics elements in future IFE facilities, where they could sustain the direct exposure to the target and the extreme levels of radiation and debris while protecting the rest of the laser chain.

However, because the refractive index of gases is generally close to 1, the maximum  index modulation achievable in gases is much smaller than what is routinely done in solids. The length of gas-based diffractive elements must be increased to compensate the small index modulation in order to reach high diffraction efficiencies, which can be challenging for practical applications. For example, an acousto-optic grating was recently demonstrated in air using an ultrasound transducer \cite{SchrodelNP23}; achieving 50\% diffraction efficiency required propagating through a total of approximately 0.5 meters of grating length (seven passes through a 7 cm grating). An alternative technique consists in launching acoustic waves in a gas via the absorption of a spatially-modulated ``imprint'' laser beam; the modulated absorption leads to localized heating of the gas, which launches an acoustic/entropy wave. This scheme, proposed by Michine and Yoneda \cite{MichineCP20}, can generate waves with much larger refractive index modulation amplitudes $\delta n$ ($\delta n \sim 10^{-5}$--10$^{-4}$, vs. $\sim 10^{-7}$ in Ref. \cite{SchrodelNP23}), which then leads to higher diffraction efficiencies in much shorter grating sizes: Ref. \cite{MichineCP20} reports a robust 96\% efficiency over a 1 cm gas optics at 10 Hz operations. The larger amplitude of the index modulation also considerably increases the bandwidth of the grating: for volumetric transmission gratings, the spectral acceptance (i.e., the maximum bandwidth that can be diffracted by the grating) for a pulse incident at the Bragg angle and assuming small incidence angle is $\Delta\lambda/\lambda\approx 2 \delta n (\Lambda/\lambda)^2$, where $\Delta\lambda$ is the bandwidth, $\lambda$ the wavelength and $\Lambda$ the grating wavelength---and the angular acceptance (maximum deviation from the Bragg angle) is $\Delta\theta\approx \delta n \Lambda/\lambda$ \cite{YehBook}. Gratings with very small bandwidths can only diffract near-perfectly collimated and narrow-band (i.e., long-pulse) beams; the very high gas density modulation achievable using photochemically-induced acoustic waves allows much higher bandwidths than using transducers, which should expand the range of applications of these optical elements.

\begin{figure}[htbp]
\includegraphics{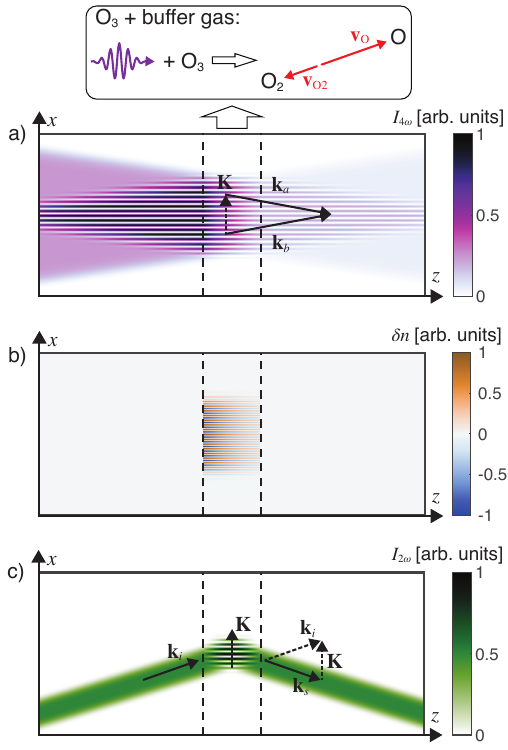}
\caption{Illustration of gas optics mechanism: a) a spatially modulated imprint laser with its wavelength $\lambda_\textsc{uv} \in$ [200 -- 300] nm gets absorbed in a gas mixture (located between the dashed lines) containing a small fraction of ozone and a buffer gas. The absorption dissociates the ozone molecules, leading the rapid heating of the surrounding gas in the ``bright'' UV fringes. b) The gas heating initiates an acoustic/entropy wave, whose density modulation has an associated refractive index modulation---thus turning the gas into a transmission grating. Using a UV laser with clean phase front and a flat-top intensity profile can in principle achieve a high-quality grating if the gas volume is well-contained within the UV modulation volume along $z$, as illustrated here. c) A high-power laser beam (with 527 nm wavelength in this illustration) incident on the gas grating at the Bragg angle can be diffracted with efficiencies near 100\%.}
\label{fig:setup}
\end{figure}

In this article, we present a comprehensive description of the physics and chemistry of acousto-optics elements in gases generated by the absorption of a modulated UV ``imprint'' beam (cf. Fig. \ref{fig:setup}). Reviewing the large body of work on the cycle of ozone in the atmosphere allowed us to identify the main chemical reactions initiated by the products of the photodissociation of ozone by UV light. We calculate the reaction enthalpies and solve the reaction rates equations coupled with the saturated absorption of the UV light;  deriving the fraction of energy available as heat for each reaction based on the relevant time-scales gives the heating rate of the gas. We then solve the linear fluid equations using the gas heating as initial conditions; we find that the isochoric heating of the gas by the chemical reactions leads to a mixed acoustic/entropy wave in the gas, where the density and temperature modulations oscillate out of phase. The density---thus refractive index---modulation from the wave  turns the gas into a transient optical grating; the index modulation is non-sinusoidal along the modulation direction $x$ and non-uniform along the longitudinal direction $z$ due to the saturated absorption of the UV light by ozone depletion. We calculate the diffraction efficiency of this gas grating and, consistent with Ref. \cite{MichineCP20}, we find that using a mixed oxygen/ozone gas with a few percent ozone fraction and a $<$100 mJ UV imprint beam over 1 cm$^2$ transverse area can generate a Bragg grating that can in principle diffract a kJ-class laser beam with nearly 100\% efficiency over less than 1 cm. {Our theoretical diffraction model is validated against 3D paraxial wave propagation simulations.} We also show that replacing oxygen by carbon dioxide as the buffer gas is expected to increase the performance of the system by almost a factor 10 (in terms of length of the gas or fraction of ozone), due to more favorable chemical reactions between the ozone photodissociation products and CO$_2$.

The rest of the article is organized as follows: in Sect. \ref{sec:DT}, we describe the saturated absorption of the UV light by ozone, and the subsequent chemical reactions leading to localized and isochoric gas heating. Section \ref{sec:acoustic} describes the dynamics of the acoustic and entropy waves initiated by the gas heating, and the resulting formation of a spatially modulated refractive index in the gas (i.e., an optical grating). Section \ref{sec:diffract} discusses the diffraction efficiency of a laser beam incident onto the grating at the Bragg angle. Finally, Sect. \ref{sec:discuss} describes how the scheme can potentially be optimized by using a different buffer gas, e.g., by using CO$_2$ instead of O$_2$.

\section{UV absorption and gas heating \label{sec:DT}}%%%%%%%%%%%%%%%%%%%%%%%%

In this section we describe how the absorption of UV light by ozone is converted into heat, both via collisional relaxation of the dissociation products (O and O$_2$) and via chemical reactions between the products and the surrounding buffer gas. The general setup, similar to Ref. \cite{MichineCP20}, is illustrated in Fig. \ref{fig:setup}. A spatially modulated UV laser (as can be generated by the interference between two beams, or passing a single beam through a Fresnel biprism) whose wavelength is in the ozone Hartley band ($\lambda \in$ [200 -- 300] nm) propagates through a gas mixture containing a small fraction of ozone and a buffer gas (e.g., O$_2$). Upon absorption, each UV photon dissociates one ozone molecule into O and O$_2$. The dissociation products have translational energies far greater than the background gas' (assuming room temperature), and will rapidly heat the surrounding gas molecules via collisions. The products also initiate exothermic chemical reactions with the gas, which can contribute significantly to the gas heating, as will be shown below.

The absorption of ultraviolet light by ozone in the Hartley band ($\lambda \in$ [200 -- 300] nm) is a well-documented process due to its importance in the absorption of UV light by the atmosphere \cite{JohnstonARPC92}. Absorption of a UV photon in the Hartley band leads to the photodissociation of an ozone molecule into O and O$_2$, following two channels:
\bea
h\nu + \text{O}_3 &\rightarrow& \text{O}_2(^1\Delta_g)+\text{O}(^1D) \, \label{eq:1D}  \\
 &\rightarrow& \text{O}_2(^3\Sigma_g)+\text{O}(^3P) \,  \label{eq:3P} .
\eea

The first channel, often referred to as the ``1D channel'' in the literature, occurs with 90\% probability and leads to the formation of O and O$_2$ in electronically-excited states, with electronic energies of 1.97 eV and 0.98 eV for O($^1$D) and O$_2$($^1\Delta_g$), respectively (cf. Table \ref{tab:photo}). The second, ``3P channel'', leads to the formation of O and O$_2$ in their ground electronic states, with a probability of 10\% \cite{SchinkeJCP10}.

\begin{table}[h]\centering
  \renewcommand{\arraystretch}{1.3}
    \begin{tabular}{m{0.07\textwidth} | m{0.09\textwidth} m{0.09\textwidth} | m{0.09\textwidth} m{0.09\textwidth}}
    \hline
      & O$_2$($^1\Delta_g$) & O($^1$D)  & O$_2$($^3\Sigma_g$)  &  O($^3$P) \\
    \hline
    $E_{el}$ & 0.98 & 1.97 & 0 & 0 \\
    $E_T$ & 0.19 & 0.37 & 0.74 & 1.47 \\
    $E_{rot}$ & 0.17 & 0 & 0.59 & 0 \\
    $E_{vib}$ & 0.23 & 0 & 1.15 & 0 \\
     \hline
    \end{tabular}
    \caption{Energy partition of the products of the photodissociation of ozone (in eV), from Refs. \cite{ThelenJCP95,DanielsJCP93}. $E_{el}$, $E_{T}$, $E_{rot}$ and $E_{vib}$ are the electronic, translational, rotational and vibrational energies.}
    \label{tab:photo}
\end{table}

Since the absorption of one UV photon leads to the dissociation of one ozone molecule, the energy balance for the photodissociation can be written as
\bea
h\nu = D_0 + E_{el} + E_{T} + E_{rot} + E_{vib} \,,
\eea
where $D_0$ = 1.05 eV is the photodissociation energy of ozone, and $E_{el}$, $E_{T}$, $E_{rot}$ and $E_{vib}$ are the electronic, translational, rotational and vibrational energies of the products, respectively (the translational and rotational energies of the ozone molecule are $\approx$ 0.04 eV for room temperature, and are thus neglected since they are typically much smaller than the energies of the products). The energy partition for the dissociation products for a UV wavelength of 248 nm ($h\nu$ = 5 eV) was measured in Refs. \cite{ThelenJCP95,DanielsJCP93} and is reported in Table \ref{tab:photo}. Approximately 4 eV of excess energy ($h\nu-D_0$) is available for the products after the 5 eV photon has dissociated the O$_3$ molecule. For the 1D photodissociation channel, approximately 3 eV go into excited electronic states of the products, leaving 1 eV available into translational, rotational and vibrational energy. The 3P channel provides 4 eV available as translational, rotational and vibrational energy.

The first (and most important) mechanism for heating the surrounding gas is the ``direct'' collisional relaxation of the photodissociation products with the gas molecules. The collision frequency of a fast particle against a background of slow particles (assumed stationary) is given by $\nu_c=\pi (r_{fast}+r_{slow})^2v_{fast}n_{bgd}$ \cite{KauzmannBook}, with $r_{fast}, r_{slow}$ the kinetic radii of the fast and slow particles, $v_{fast}$ the fast particle velocity and $n_{bgd}$ the density of the background gas. The kinetic collision diameters of O and O$_2$ are 2.5 \AA\, \cite{TaniguchiJPCA00} and 3.46 \AA, respectively. We obtain collision times ($\tau_c=1/\nu_c$) for the fast dissociation products against the background oxygen gas of 100 ps and 68 ps for O$_2(^1\Delta_g)$ and O($^1$D), and 51 ps and 34 ps for O$_2(^3\Sigma_g)$ and O($^3$P). The mean free paths $l_c=v_{fast}\tau_c=1/[\pi n_{bgd}(r_{fast}+r_{slow})]$ are 0.11 $\mu$m and 0.14 $\mu$m for O$_2$ and O (independent of the velocity for hard sphere collisions, thus independent of the dissociation channel).

\begin{table*}[t]
\begin{tabular}{ | m{0.05\textwidth} | m{0.35\textwidth} | m{0.15\textwidth} | m{0.12\textwidth} | m{0.1\textwidth} | m{0.08\textwidth} | }
\hline
\rule{0pt}{15pt}\myb{No.} & \myb{Reaction} & \myb{Rate [cm$^3$/s]} & \myb{$-\Delta_r$H$^0$ [eV]} & \myb{E$_{heat}$ [eV]} & \myb{Ref.}    \\
\hline
\hline
0a & $h\nu$+O$_3\rightarrow$ O$_2$($^1\Delta_g$)+O($^1$D) & $k_{0a}$=0.9$\times$3.3e-7 & 0.96 & 0.73 & \\
0b & $h\nu$+O$_3\rightarrow$ O$_2$+O & $k_{0b}$=0.1$\times$3.3e-7 & 3.95 & 2.8 & \\
1a & O($^1$D)+O$_2\rightarrow$ O+O$_2$($^1\Sigma_g$) & $k_{1a}$=0.8$\times$3.95e-11 & 0.34 & 0.29 & \cite{BurkholderNASA19} \\
1b & O($^1$D)+O$_2\rightarrow$ O+O$_2$($^1\Delta_g$) & $k_{1b}$=0.2$\times$3.95e-11 & 0.99 & 0.84 & \cite{BurkholderNASA19} \\
2a & O($^1$D)+O$_3\rightarrow$ 2O$_2$ & $k_{2a}$=1.2e-10 & 6.0 & 4.3 & \cite{BurkholderNASA19} \\
2b & O($^1$D)+O$_3\rightarrow$ O$_2$+2O & $k_{2b}$=1.2e-10 & 0.9 & 0.81 & \cite{BurkholderNASA19} \\
3a & O$_2$($^1\Sigma_g$)+O$_3\rightarrow$ O+2O$_2$ & $k_{3a}$=1.2e-11 & 0.53 & 0.42 & \cite{GreenJPCA00} \\
3b & O$_2$($^1\Sigma_g$)+O$_3\rightarrow$ O$_3$+O$_2$ & $k_{3b}$=1e-11 & 1.63 & 1.16 & \cite{GreenJPCA00} \\
\hline
\end{tabular}
\caption{Summary of all the reactions contributing to the heating of the gas over the few ns timescales relevant to gas optics. $\Delta_r$H$^0$ is the reaction enthalpy, and E$_{heat}$ is the sum of translational and rotational energies of the products, which constitutes the energy available to heat the surrounding gas molecules over the relevant time scales for each reaction.}
\label{tab:reactions}
\end{table*}

The mean free paths and collision times are thus much shorter than the UV modulation period ($\Lambda\sim$ few tens of $\mu$m) and the temporal scale of the subsequent acoustic waves ($\tau_s=\Lambda/c_s\sim$100 ns, with $c_s$ the sound speed in the gas---cf. Sect. \ref{sec:acoustic}). We can therefore assume that all the translational and rotational energy of the dissociation products will contribute to heating the gas locally and instantaneously (i.e., isochorically). On the other hand, the dissipation of the vibrational energy is typically much slower and is not expected to contribute to gas heating at our time scales; the relaxation rates of O$_2(^3\Sigma_g$) were measured to be on the order of $\mu$s$^{-1}$ at most \cite{RogaskiFD95}.

While the direct collisional relaxation of the translational and rotational energies of the products will constitute the main source of heating for the duration of the UV pulse (10 ns in Ref. \cite{MichineCP20} and in this article), the chemical reactions between the photodissociation products and the surrounding buffer gas, including the ``quenching'' of the electronic energy for the 1D channel, will also be a significant contributor. Table \ref{tab:reactions} summarizes all the reactions contributing to the overall heating and energy balance over a typical 10 ns time scale. Other processes, like the radiative de-excitation of the excited electronic states or termolecular (three reactants) reactions, are too slow to contribute to the localized gas heating initiating the acoustic wave. In particular, the regeneration of ozone via O+2O$_2\rightarrow$O$_2$+O$_3$ has a reaction rate of $\approx 6\times 10^{-34}$ cm$^6$/s, which occurs over $\sim \mu$s timescale at room temperature in our conditions. In this table and throughout the rest of the paper, the notations O, O$_2$ and O$_3$ without specification of the electronic state configuration will usually refer to the electronic ground state of the atoms or molecules.

All the reactions in Table \ref{tab:reactions} are exothermic, with the reaction enthalpy $\Delta_rH^0=\Sigma \Delta_fH^0$(products) $-\Sigma \Delta_fH^0$(reactants). The formation enthalpies $\Delta_fH^0$ for the atoms and molecules considered here are given in Table \ref{tab:DfH}. For consistency with the other reactions we defined the reaction rate for the photodissociation process as $k_0=k_{0a}+k_{0b}=\sigma c$, with $\sigma=1.1\times 10^{-17}$ cm$^2$ the ozone absorption cross-section in the center of the Hartley band and $c$ the speed of light.

\begin{table}[h]\centering
  \renewcommand{\arraystretch}{1.3}
    \begin{tabular}{m{0.2\textwidth} m{0.2\textwidth}}
    \hline
    \myb{O} ($\Delta_fH^0=$ 2.58) & \myb{O$_2$} ($\Delta_fH^0=$ 0) \\
    \hline
    O($^3$P):  0 & O$_2$($^3\Sigma_g$):  0  \\
    O($^1$D): 1.97 & O$_2$($^1\Delta_g$):  0.98   \\
     & O$_2$($^1\Sigma_g$):  1.63   \\
     \hline
    \end{tabular}
    \caption{Electronic energies of the excited levels of O and O$_2$ considered in this paper (all numbers are expressed in eV). The formation enthalpies $\Delta_fH^0$ are for the ground electronic states; for ozone, $\Delta_fH^0=$1.48 eV.}
    \label{tab:DfH}
\end{table}

The reaction energy available to heat the gas, $E_{heat}$ in Table \ref{tab:reactions}, is the sum of the translational and rotational energies. In the absence of measurements of the energy partition of the reaction products (between translational, rotational and vibrational modes) except for the photodissociation process (reaction 0, cf. Table \ref{tab:photo}), we make the assumption of equilibrium equipartition in order to estimate $E_{heat}$ for all the other reactions, with the vibrational modes activated. The monoatomic O molecule has 3 modes for translational energy, O$_2$ (linear molecule with 2 atoms) has 3, 2 and 2 modes for translational, rotational and vibrational energies, and O$_3$ (three atoms, non-linear) has 3, 3, and 6 modes for translational, rotational and vibrational energies. We then use conservation of momentum between the reaction products (neglecting the initial center of mass kinetic energy of the reactants, since the products are generated with velocities much higher than the background gas thermal velocity) to calculate the ratio of translational energy between the products, which ultimately gives the ``weight'' of each energy mode for each of the products. We can then calculate $E_{heat}$ by taking the available energy from the reaction enthalpy and removing the vibrational energy of the polyatomic molecules. For example, for reaction 1a, O($^1$D)+O$_2\rightarrow$ O+O$_2$($^1\Sigma_g$):
\bi
\item conservation of momentum for the products, $m_\text{O}v_\text{O}=m_\text{O2}v_\text{O2}$, yields $E_T$(O) = 2$E_T$(O$_2$);
\item equilibrium energy equipartition for O$_2$, with 3, 2 and 2 modes for translational, rotational and vibrational energies, respectively, leads to $E_{rot}$(O$_2$) = $E_{vib}$(O$_2$) = $\frac 23 E_T$(O$_2$), and thus to the relative energy ``weights'' [6,3,2,2] for [$E_T$(O), $E_T$(O$_2$), $E_{rot}$(O$_2$), $E_{vib}$(O$_2$)] with $E_T$(O)+$E_T$(O$_2$)+$E_{rot}$(O$_2$)+$E_{vib}$(O$_2$) = $-\Delta_rH^0$;
\item thus, $E_{heat,1a}$ = $-\Delta_rH^0-E_{vib}$(O$_2$) = $-\frac{11}{13}\Delta_rH^0$ = 0.29 eV.
\ei

For reactions with three products, we assume equipartition for the translational energies of the products \cite{BaerJCP82}, such that the ratio of translational energy of one product $E_{T,i}$ to the total for all the products, $E_{T,tot}$, is given by
\bea
\frac{E_{T,i}}{E_{T,tot}} = \frac{1-m_i/M}{N_p-1} \,,
\eea
where $m_i$ the mass of the product $i$, $M$ is the sum of the masses of all the products, and $N_p$ is the number of products. For example, for reaction 2b, O($^1$D)+O$_3\rightarrow$ O$_2$+2O, we obtain the relative weights [2,3,3] for [$E_T$(O$_2$), $E_T$(O), $E_T$(O)].

In reality the products are not at equilibrium; however, assuming equilibrium equipartition is not expected to introduce a large error on $E_{heat}$ since the only difference will be on the fraction of energy contained in the vibrational modes. For example, assuming equipartition for reaction 0 leads to a $<$10\% error on $E_{heat}$ compared to the values in Table \ref{tab:photo}, based on actual measurements.

To estimate the total gas heating, we must calculate the evolution of the different species and reactions for the duration of the UV imprint beam, coupled with the absorption of the UV laser (which is saturated due to the depletion of ozone via photodissociation). We solve the following system of equations numerically, as a function of time and propagation direction $z$ for an arbitrary $x$ (we assume that the UV intensity modulation is along $x$, cf. Fig \ref{fig:setup}):
\bea
\dt [\Oz ] &=& -(k_{0a}+k_{0b})[h\nu][\Oz]  \nonumber \\
&&- (k_{2a}+k_{2b})[\Oz][^1\text{D}] \nonumber \\
&&- (k_{3a}+k_{3b})[^1\Sigma_g][\Oz]  \, ,\,\,\,\,  \label{eq:CME1} \\
\dt [^1\text{D}] &=& k_{0a}[h\nu][\Oz] \nonumber \\
&& - (k_{1a}+k_{1b})[^1\text{D}][\text{O}_2] \nonumber \\
&& - (k_{2a}+k_{2b})[^1\text{D}][\Oz]  \, , \label{eq:CME2} \\
\dt [^1\Delta_g] &=& k_{0a} [h\nu][\Oz] + k_{1b}[^1\text{D}][\text{O}_2]  \,,  \label{eq:CME3} \\
\dt [^1\Sigma_g] &=& k_{1a}[^1\text{D}][\text{O}_2] \nonumber \\
&&- (k_{3a}+k_{3b})[^1\Sigma_g][\Oz] \,,  \label{eq:CME4} \\
(\dt+c\dz) [h\nu] &=& -c\sigma[h\nu][\Oz] \,. \label{eq:CME5}
\eea

The square brackets denote densities; we also omit the molecule to simplify the notations, e.g., [$^1$D] refers to the density of O($^1$D), etc. Upon solving Eqs. \eqref{eq:CME1}--\eqref{eq:CME5}, we obtain the spatio-temporal ($z,t$) profiles of densities for all the molecules, as well as the photon density $[h\nu]=I/ch\nu$ with $I$ the UV laser intensity and $\nu$ its photon frequency ($h$ is Plank's constant).

\begin{figure}[htbp]
\includegraphics{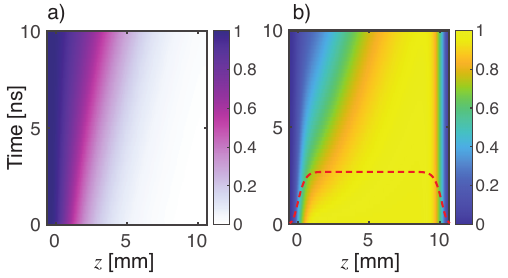}
\caption{a) Laser intensity $I(z,t)/I_0$, and b) density of ozone ${[\Oz]}(z,t)/{[\Oz]}_0$ as a function of time and propagation direction $z$, for a 10 ns square UV pulse and an initial laser intensity $I_0=1.2\times 10^7$ W/cm$^2$ (fluence $F_0=120$ mJ/cm$^2$). The gas mixture contains 2\% of ozone with oxygen as the buffer gas, at atmospheric conditions (${[\Oz]}_0=5\times 10^{17}$ cm$^{-3}$). The red dashed lineout in b) represents the spatial profile of the initial ozone density vs. $z$ (in arbitrary units).}
\label{fig:Inzt}
\end{figure}

Figure \ref{fig:Inzt} shows the laser intensity profile $I(z,t)$ and the ozone density ${[\Oz]}(z,t)$ for typical experimental conditions. The laser propagates towards $z>0$ and has a square temporal profile with 10 ns duration, and the gas volume is assumed to have a spatial density profile for the ozone along $z$ following a super-Gaussian profile of order 8 and full width at half maximum of 1 cm (red lineout in Fig. \ref{fig:Inzt}b---a well-contained gas density is easy to realize at ambient pressure using a flow tube \cite{MichineCP20}). The UV modulation is assumed to have a contrast of one and fill a larger volume than the size of the diffracted beam, so that the grating can be assumed independent of $y$, with a sharp entrance boundary and $\Lambda$ constant throughout. After 10 ns, the laser penetrates further into the gas (Fig. \ref{fig:Inzt}a) due to the depletion of ozone (Fig. \ref{fig:Inzt}b) which saturates the absorption.

Note that the laser intensity and ozone density can be very well approximated analytically, by noticing that the ozone depletion is largely dominated by the laser absorption in Eq. \eqref{eq:CME1} and that the laser transit time through the length of the gas (33 ps for 1 cm of gas) is typically much shorter than the UV pulse duration, which means that the time derivative in Eq. \eqref{eq:CME5} can be neglected. The space-time evolution of the UV intensity and ozone density is therefore dominated by only two coupled equations, $\partial_t [\text{O}_3]=-k_0[h\nu][\text{O}_3]$ and $\partial_z[h\nu]=-\sigma[h\nu][\text{O}_3]$, with $[\text{O}_3]$ and $[h\nu]$ functions of $z$ and $t$ and $k_0=k_{0a}+k_{0b}$; these coupled equations can be solved analytically---for example, they have the same form as the equations describing nonlinear crossed-beam energy transfer in two spatial dimensions. Following the same derivation described in Ref. \cite{MichelBook} (Sect. 10.1.3), we obtain the space and time dependent solutions for the saturated absorption of the UV laser and ozone density depletion:
\bea
I(z,t) &=& I_0\frac{e^{t/t_{dep}}}{e^{z/z_{abs}}-1+e^{t/t_{dep}}} \,, \\
{[\Oz]}(z,t) &=& {[\Oz]}_0\frac{e^{z/z_{abs}}}{e^{z/z_{abs}}-1+e^{t/t_{dep}}} \,, 
\eea
where $I_0=I(0,t)$ and ${[\Oz]}_0={[\Oz]}(z,0)$ are the vacuum UV intensity and the initial ozone density, and $z_{abs}=1/\sigma{[\Oz]}_0$ and $t_{dep}=h\nu/\sigma I_0$ are the linear UV absorption scale-length and ozone depletion time-scale.

We can then calculate the total heating rate by summing up the contributions from all the reactions in Table \ref{tab:reactions}, according to
\bea
\dt U_h &=& (0.73k_{0a}+2.8k_{0b})[h\nu][\Oz]  \nonumber \\
&& + (0.29k_{1a}+0.84k_{1b})[^1\text{D}][\text{O}_2]   \nonumber \\
&& + (4.3k_{2a}+0.81k_{2b}) [^1\text{D}][\Oz] \nonumber \\
&&  + (0.42k_{3a}+1.16k_{3b})[^1\Sigma_g][\Oz]  \,, \,\,
\eea
where $U_h$ is the energy density in eV/cm$^3$ (function of $z$ and $t$) that can be converted into heat. This rate accounts for both the ``direct'' heating from collisional relaxation of the photodissociation products discussed above (reactions 0a and 0b), as well as all the other chemical reactions between the products and the surrounding gas. 

The local temperature increase over the duration of the UV pulse is then calculated as
\bea
\Delta T(x,z) = \frac{U_h(x,z)}{{[\text{O}_2]}c_v(\text{O}_2)} \,,  \label{eq:DT}
\eea
where $U_h(x,z)=\int_0^\tau \partial_t U_h(x,z,t) dt$ is the total deposited energy density available for heating the gas at the end of the UV pulse (of duration $\tau$), ${[\text{O}_2]}$ is the density of the background oxygen and $c_v$(O$_2$) its  isochoric heat capacity. We assumed that the ozone fraction was small enough to neglect its contribution (and the contribution of the other reactions' products) to the heat capacity of the gas. The initial modulated UV intensity profile at $z=0$ is assumed to be of the form $I(x,z=0)=I_0[1+\cos(Kx)]$ for $0<t<\tau$, with $K=2\pi/\Lambda$ and $\Lambda$ the modulation wavelength (i.e., the grating spatial period).

\begin{figure}[htbp]
\includegraphics{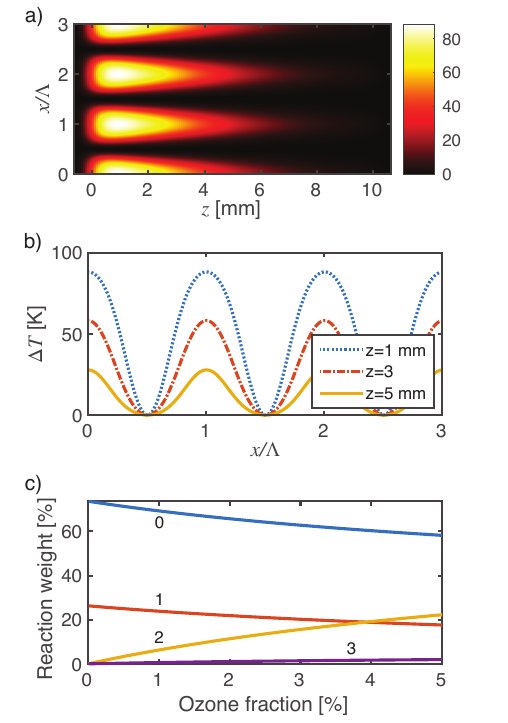}
\caption{a) Temperature increase $\Delta T$ [K] for the background O$_2$ gas at the end of the 10 ns UV pulse, for an average UV fluence of 60 mJ/cm$^2$ (average intensity $I_0=6\times 10^6$ W/cm$^2$) and an initial fraction of ozone of 2\%. b) Temperature profile lineouts at $z$ = 1, 3 and 5 mm. c) Contributions of the reactions 0, 1, 2 and 3 from Table \ref{tab:reactions} to the total gas heating (integrated in space and over the duration of the UV pulse) as a function of ozone fraction.}
\label{fig:DTxz}
\end{figure}

A typical heating profile $\Delta T(x,z)$ is shown in Fig. \ref{fig:DTxz}, at the end of a 10 ns UV pulse with average fluence $F_0=\tau I_0=$ 60 mJ/cm$^2$  and a 2\% ozone concentration. Figure \ref{fig:DTxz}b shows lineouts at $z$ = 1, 3 and 5 mm. The lineouts show evidence of saturation of the gas heating due to ozone depletion: the transverse profile at the beginning of the UV propagation ($z=$ 1 mm), where heating is maximum (because the UV light has not been fully absorbed yet) is non-sinusoidal; whereas the profile after further propagation becomes sinusoidal again, like the UV transverse intensity profile. Note that the ozone depletion shown in Fig. \ref{fig:Inzt} for a fluence of $1.2\times 10^7$ W/cm$^2$ corresponds to the $x=0$ location (modulo $\lambda$) on Fig. \ref{fig:DTxz}, where $I(x,z=0)=I_0[1+\cos(Kx)]$ with $I_0=6\times 10^6$ W/cm$^2$. Figure \ref{fig:DTxz}c shows the contributions of the different reactions from Table \ref{tab:reactions} to the total heating under the same UV fluence conditions. The direct heating by collisional relaxation of the photodissociation products (reaction 0) dominates, but the other reactions also contribute significantly; in particular, the contribution of reactions 2 and 3 increases with higher ozone fractions, since ozone is a reactant for these reactions.

The modulated gas heating by the UV beam then sets the initial conditions for the acoustic/entropy wave, as the resulting pressure modulation launches hydrodynamic gas motion upon release, as discussed in the next section.

\section{Acoustic/entropy wave\label{sec:acoustic}}%%%%%%%%%%%%%%%%%%%%%%%%

To investigate the gas response to the modulated heating, we describe the gas by its fluid quantities $\rho$, $v$ and $p$ (mass density, velocity and pressure). Because variations along $z$ are dictated by the UV absorption profile, which is on the order of the linear absorption length $z_{abs}=1/\sigma{[\Oz]}_0$, they typically occur over distances much longer than $\Lambda$ (the characteristic spatial scale along $x$); for example, for 2\% ozone fraction we have $z_{abs}=$ 1.8 mm, whereas $\Lambda$ is typically on the order of a few tens of microns. We can therefore assume that the hydrodynamics is 1D along $x$. The following analysis is assumed to be for a given $z$, with the initial amplitude of the temperature modulation provided by the analysis from the previous section. We start with the Euler fluid equations:
\bea
\dt \rho + \dx(\rho v) &=& 0 \,, \label{eq:Euler1} \\
\dt v + (v\dx)v &=& -\frac{\dx p}{\rho} \, . \label{eq:Euler2}
\eea

We linearize the fluid quantities, $\rho=\rho_0+\rho_1$ etc., and expand the Euler equations to first order:
\bea
\dt \rho_1 + \rho_0\dx v_1 &=& 0 \, , \label{eq:E1} \\
\dt v_1 &=& -\frac{\dx p_1}{\rho_0} \,.  \label{eq:E2}
\eea

The isochoric gas heating resulting from the UV laser absorption creates an initial temperature modulation in the gas while the density remains constant over the duration of the UV pulse. Once the heating is done, the gas is set in motion adiabatically; its response to the initial temperature disturbance follows $\rho\propto p^{1/\gamma}$, with $\gamma$ the adiabatic index. Inserting into Eq. \eqref{eq:E1} gives $\dt \rho_1 = (\rho_0/\gamma p_0)\dt p_1$. Taking the time derivative of  Eq. \eqref{eq:E1}, spatial derivative of Eq. \eqref{eq:E2} and combining the two gives the usual wave equation for the pressure perturbation:
\bea
(\dt^2-c_s^2\dx^2)p_1 = 0 \,,  \label{eq:wave}
\eea
where $c_s=(\gamma p_0/\rho_0)^{1/2}$ is the sound speed. Using the d'Alembert formula with the initial pressure perturbation $p_1(x,t=0)=R_s\rho_0T_1=R_s\rho_0\Delta T(x)$, with $R_s$ the specific gas constant and $T_1=\Delta T(x)$ the modulated temperature increase at the end of the UV pulse calculated in the previous section, immediately gives:
\bea
\frac{p_1(x,t)}{p_0} = \frac 12 \left[ \frac{\Delta T(x+c_st)}{T_0} + \frac{\Delta T(x-c_st)}{T_0} \right] \,, \label{eq:p1}
\eea
where $T_0$ is the initial gas temperature (the 10 ns duration of the UV pulse is treated as an initial impulse compared to the slower time scale of the wave). The pressure perturbation forms the expected standing wave, as the sum of two counter-propagating waves propagating at $c_s$ towards $\pm x$.

Next we solve for the other fluid quantities. We first integrate Eq. \eqref{eq:E2} vs. time to get $v_1=-(1/\rho_0)\int \dx p_1 dt$ (with $v_1(t=0)=0$). Inserting into Eq. \eqref{eq:E1} and using Eq. \eqref{eq:wave} gives $\rho_1=p_1/c_s^2+f(x)$, where $f$ is a constant of integration depending on the initial conditions. Since $\rho_1(t=0)=0$, we have $f(x)=-p_1(x,t=0)/c_s^2$, which leads to the solution for the density perturbation:
\bea
\frac{\rho_1(x,t)}{\rho_0} = \frac {1}{2\gamma} \left[ \frac{\Delta T(x+c_st)}{T_0} + \frac{\Delta T(x-c_st)}{T_0} \right ] \nonumber \\
- \frac 1\gamma \frac{\Delta T(x)}{T_0} \,.   \label{eq:n1}
\eea

Finally, we can derive the temperature perturbation $T_1$ via $p_1=R_s\rho_0T_1+R_s\rho_1T_0$, leading to
\bea
\frac{T_1(x,t)}{T_0} = \frac{\gamma-1}{2\gamma}\left[ \frac{\Delta T(x+c_st)}{T_0} + \frac{\Delta T(x-c_st)}{T_0}\right] \nonumber \\
+ \frac 1\gamma \frac{\Delta T(x)}{T_0} \,. \label{eq:T1}
\eea

These expressions for the density and temperature modulations follow a typical acoustic (standing-)wave behavior, except for the presence of the non-propagating term $\Delta T(x)$ in the equations. This disturbance is associated with the initial conditions, and corresponds to an ``entropy mode''. To shed light on its meaning, we consider the entropy $s=c_v\ln(p)-c_p\ln(\rho)+cst.$ and linearize it to first order like the other fluid quantities, $s=s_0+s_1$; the first order perturbation is:
\bea
s_1 &=& c_v\left[ \frac{p_1}{p_0}-\gamma\frac{\rho_1}{\rho_0} \right] \label{eq:s1temp} \\
  &=& c_v\frac{\Delta T(x)}{T_0} \,, \label{eq:s1}
\eea
where we used the definition of the adiabatic index $\gamma=c_p/c_v$ and substituted the expressions from Eqs. \eqref{eq:p1}--\eqref{eq:n1} in Eq. \eqref{eq:s1temp}.

We recognize the non-propagating term appearing in the density and temperature equations, Eqs. \eqref{eq:n1}--\eqref{eq:T1}. Physically, this comes from the initial heating of the gas by the chemical reactions: because the heating is isochoric, the entropy of the gas increases with the temperature, leading to an initial entropy modulation along $x$ at time 0, $s_1(x)= c_v\Delta T(x)/T_0$. Because the initial temperature modulation is positive everywhere (the gas is heated by the reactions), the entropy also increases everywhere. After the heating has taken place, the adiabatic equation of state for the evolution of the gas implies that the entropy must stay constant, $\partial_t s=0$: therefore, the initial entropy modulation introduced by the rapid chemical heating remains throughout the subsequent evolution of the gas, as a constant, non-propagating perturbation of density and temperature. Note that the entropy mode does not impact the pressure: indeed, defining $\rho_{1s}/\rho_0=-\Delta T(x)/\gamma T_0$ and $T_{1s}/T_0=\Delta T(x)/\gamma T_0$ the density and temperature perturbations associated with the entropy mode, and decomposing the pressure perturbation into an acoustic and entropy components, $p=R_s\rho T=R_s(\rho_0+\rho_{1ac}+\rho_{1s})(T_0+T_{1ac}+T_{1s})$, we see that the pressure perturbation $p_{1s}$ associated with the entropy mode is equal to $p_{1s}=R_s(\rho_0T_{1s}+T_0\rho_{1s})=0$.

To summarize: the resulting wave, as described by Eqs. \eqref{eq:p1}--\eqref{eq:s1}, is a mixed acoustic/entropy wave, comprised of a standing acoustic wave (sum of two counter-propagating waves) and a constant, non-propagating entropy mode $s_1(x)$ with no associated pressure perturbation. If the initial heating of the gas had been adiabatic instead of isochoric, there would have been no initial entropy modulations, and only the standing acoustic wave would have been present. In a system of finite size along $x$, the two counter-propagating acoustic waves should eventually exit the system, leaving the steady-state entropy mode behind---until the slow diffusion in the gas eventually damps it out \cite{MichinePF24}. Because the collision mean free path is much smaller than the physical scale of the wave (i.e., its wavelength $\Lambda$), diffusion should occur over time scales much greater than the ones we are considering here. Our analysis is consistent with Ref. \cite{CummingsAO95} when neglecting spatial envelope effects and diffusion (cf. their Eq. (18)). In that paper, the entropy mode was described as a ``thermal grating'' or ``thermon''; our different nomenclature aims to avoid confusion with the ``thermal gratings'' introduced in Ref. \cite{EichlerJAP73}, where the refractive index modulation is created by the temperature-dependence of the index, which is different from our configuration---and could not achieve the type of index modulation described here.

An example of hydrodynamic evolution of the gas is shown in Fig. \ref{fig:wave}, for the initial temperature modulation calculated in the previous section, taken from Fig. \ref{fig:DTxz}b at $z=1$ mm. {The presence of the entropy mode leads to out-of-phase oscillations between the temperature and density modulations. This is different from a standard standing acoustic wave, with adiabatic initial conditions for the pressure perturbation, where density and temperature oscillate in phase.

\begin{figure}[htbp]
\includegraphics{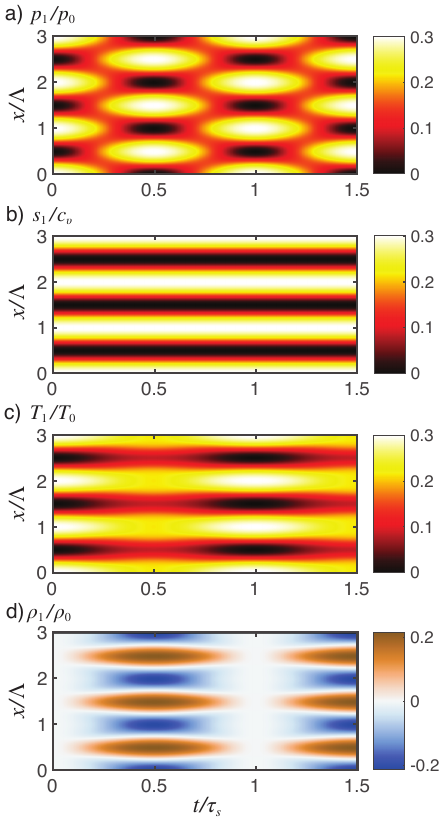}
\caption{{Acoustic/entropy wave dynamics initiated by an initial gas heating $\Delta T(x)$ taken from Fig. \ref{fig:DTxz}b at $z=1$ mm, with $\Delta T_{max}=$ 88 K, i.e., $\Delta T_{max}/T_0=$ 0.3 for room temperature: a) pressure perturbation $p_1/p_0$, b) entropy, $s_1/c_v$, c) temperature, $T_1/T_0$, and d) density, $\rho_1/\rho_0$. Time is normalized to the acoustic period $\tau_s=\Lambda/c_s$ and space to the modulation wavelength $\Lambda$.}}
\label{fig:wave}
\end{figure}

The expressions for the fluid quantities, Eqs. \eqref{eq:p1}--\eqref{eq:T1}, show that for an initial temperature disturbance $\Delta T(x)$ with spatial period $\Lambda$, the maximum amplitude of the density modulation is reached at $t=\tau_s/2 + j\tau_s (j\in\mathbb{N})$ with $\tau_s=\Lambda/c_s$ (cf. Fig. \ref{fig:wave}d), with
\bea
\frac{\rho_1(x,\tau_s/2)}{\rho_0} = \frac{1}{\gamma} \frac{\Delta T(x+\Lambda /2,z)-\Delta T(x,z)}{T_0} \, .
\eea

For a sinusoidal temperature modulation, with $\Delta T(x+\Lambda/2)=-\Delta T(x)$, we obtain $\rho_{1,max}/\rho_0=(2/\gamma)T_{1,max}/T_0$. The coefficient $2/\gamma$ is due to the presence of the entropy mode: for a pure acoustic mode, the proportionality coefficient would be $1/(\gamma-1)$.} For oxygen, with $\gamma=7/5$, the peak density modulation is therefore equal to 10/7 times the initial peak temperature modulation introduced by the UV laser. The density modulation amplitudes generated by this technique can be up to tens of percents---e.g., it is $\pm$ 21\% in Fig. \ref{fig:wave}, with $\Delta T_{max}/T_0=0.3$. This very large amplitude is consistent with the measurements from Ref. \cite{MichineCP20}; it corresponds to an equivalent sound level of about 180 dB, far beyond what piezoelectric transducers can deliver.

\begin{figure}[htbp]
\includegraphics{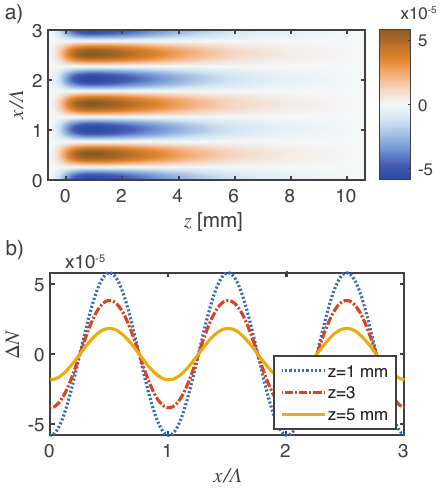}
\caption{Refractive index modulation (i.e., optical grating) created at $t=\tau_s/2$ with the same conditions as the previous figures (60 mJ/cm$^2$ of average UV fluence and 2\% ozone concentration). a) Index modulation $\Delta n(x,z)$; b) lineouts along $x$ taken at $z=$ 1, 3 and 5 mm.}
\label{fig:dNxz}
\end{figure}

Finally, since the index of refraction $n$ of the gas obeys $n-1\propto \rho$ (with $n-1\ll 1$), we obtain the refractive index modulation at $t=\tau_s/2$ as a function of the initial temperature modulation from Eq. \eqref{eq:n1}:{
\bea
\Delta n(x,z,\frac{\tau_s}{2}) &=& (n_0-1)\frac{\rho_1(x,z,\tau_s/2)}{\rho_0} \\
  &=& \frac{n_0-1}{\gamma} \frac{\Delta T(x+\Lambda /2,z)-\Delta T(x,z)}{T_0} \,, \nonumber \\ \label{eq:DN}
\eea
where $\Delta T(x,z)$ is the temperature modulation derived in the previous section, Eq. \eqref{eq:DT}, and $n_0$ the background index of the gas. Figure \ref{fig:dNxz} shows the spatial profile of the index modulation $\Delta n(x,z)$ at $t=\tau_s/2$ created with the same conditions as the previous figures. The maximum (peak-to-valley) index modulation amplitude reaches $\Delta n_{max}\approx 10^{-4}$ under these conditions. Like for the temperature profile in Fig. \ref{fig:DTxz}b, the transverse refractive index profile by the entrance of the gas cell ($z$ = 1 mm), where the UV absorption---and thus the gas heating---is saturated, is generally non-sinusoidal. Here, however, even at $\pm$20\% amplitude the density modulations remain largely sinusoidal.}

\section{Diffraction off the index modulation\label{sec:diffract}}%%%%%%%%%%%%%%%%%%%%%%%%

{We now wish to calculate the diffraction of a laser beam off the optical grating associated with the index modulation $\Delta n(x,z)$ from acoustic/entropy waves derived in the previous section. In general, light incident on a grating (or any structure with a periodic index modulation) along the wave vector $\myb k_i$ can be diffracted into multiple orders, along the wave-vectors $\myb k_m=\myb k_i-m\myb K$, with $m\in \mathbb N$ and $|\myb k_i|=|\myb k_m|=k$ ($\myb K$ is the grating wave vector, with $|\myb K|=K=2\pi/\Lambda$), allowing constructive interference of the grating's scattering centers (cf. Fig. \ref{fig:kdiag}). Limiting diffraction into a single order, in order to use the grating as a mirror, can be achieved by tailoring the shape of the index modulation for ``thin'' gratings. On the other hand, for ``thick'' or volumetric gratings, under certain conditions the scattered energy can be concentrated into the first diffraction order only ($m=1$). The criteria for defining thin vs. volumetric gratings (or ``Raman-Nath'' vs. ``Bragg'' diffraction \cite{BoydBook}) is often described via the parameter $Q=\lambda L/\Lambda^2$, where $\lambda$ is the diffracted beam's wavelength, $L$ the grating thickness (along $z$ in our geometry) and $\Lambda$ the grating wavelength (along $x$) \cite{BoydBook,KleinIEEE67}: for $Q\gg 1$, a beam incident at the Bragg angle onto the grating will diffract almost all of its energy in the first order only, whereas for $Q\le 1$, the higher-order modes will contain significant energy. A more accurate parameter, taking into account the diffraction efficiency via the index modulation amplitude $n_1$, was later introduced by Moharam and  Young \cite{MoharamAO78} as
\bea
\rho = \frac{\lambda^2}{\Lambda^2 n_0n_1} \,, \label{eq:rhodiff}
\eea
where $n_0$ and $n_1$ are the background index and the grating's index modulation amplitude, respectively. For example, for a gas optics volumetric grating like the one calculated in the previous section and shown in Fig. \ref{fig:dNxz}, we have $n_1\approx 5\times 10^{-5}$, $n_0\approx 1$, so $\rho \gg 1$ (Bragg diffraction regime) requires $\Lambda \ll 140 \lambda$. In other words: for a given amplitude of the index modulation, the grating wavelength should not be too large in order to avoid getting significant energy diffracted into higher-order modes. %This condition might require some design compromise against the diffracted beam pulse duration: indeed, reducing $\Lambda$ leads to a shorter acoustic period $\tau_s=\Lambda/c_s$, which decreases the duration for which the index modulation is maximum near $t=\tau_s/2$ (cf. Fig. \ref{fig:wave}c). For example, for 532 nm diffracted light, with $n_1\approx 5\times 10^{-5}$, diffraction in the Bragg regime requires $\Lambda \le$ 75 $\mu$m, which in turns gives the acoustic period $\tau_s$ = 227 ns---meaning that for this level of index modulation, efficient diffraction into the first order only would only be possible for pulses of durations shorter than (approximately) 100 ns.

\begin{figure}[htbp]
\includegraphics{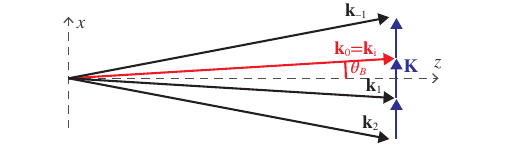}
\caption{Geometry of the diffraction of a beam with wave vector $\myb k_i$ incident onto a grating with wave vector $\myb K$ at the Bragg angle $\theta_B=\text{asin}(K/2k)=\text{asin}(\lambda/2\Lambda)$. Diffraction off the grating occurs along the directions $\myb k_m=\myb k_i-m\myb K, m\in \mathbb N$, where $m$ is the diffraction order. If $\rho\gg 1$, with $\rho$ defined in Eq. \eqref{eq:rhodiff}, diffraction occurs in the Bragg regime and most of the energy can be contained in the single $m=1$ diffraction mode.}
\label{fig:kdiag}
\end{figure}

Estimating the diffraction efficiency of volumetric gratings can be achieved with great accuracy using coupled mode theory \cite{KogelnikBSTJ69}. The calculation neglects high-order diffraction modes, which is accurate as long as $\rho\gg 1$ in Eq. \eqref{eq:rhodiff}, and assumes that the incident and first-order diffracted waves have slowly-varying amplitudes, leading to the following expression for the total electric field:
\bea
E(x,y,z,t) = \frac 12 E_i(z,t)e^{i\psi_i} + \frac 12 E_1(z,t)e^{i\psi_1} + c.c. \,,
\eea
where $c.c.$ denotes the complex conjugate, $E_i$ and $E_1$ are the slowly-varying envelopes of the incident ($m=0$) and $m=1$ diffraction modes (only dependent on $z$ and slowly varying in time compared to the laser frequency, since the system is assumed infinite along $y$ and translationally invariant along $x$), and the rapid phases are $\psi_{i,1}=\myb k_{i,1}\cdot \myb r-\omega t$. Inserting into the wave equation $(\nabla^2+n^2\omega^2/c^2)E=0$ with $n=n_0+n_1(z)\cos(Kx)$, $n_1\ll n_0$, taking a paraxial approximation and collecting the terms $\propto e^{i\psi_i}$ and $\propto e^{i\psi_1}$ leads to the simple system of coupled equations,
\bea
\partial_z E_i &=& i\frac{k}{2}\frac{n_1(z)}{n_0}E_1 \\
\partial_z E_1 &=& i\frac{k}{2}\frac{n_1(z)}{n_0}E_i \,.
\eea
Here we assumed perfect phase-matching, i.e. incidence at exactly the Bragg angle, $\myb k_i-\myb k_1=\myb K$. Introducing the variables $\zeta=kn_1/2n_0$ and $u=\int_0^z \zeta(v)dv$ (method of the grating integral, cf. \cite{YehBook}) transforms the coupled equations above into
\bea
E_i = E_i(0)\cos(u) \,, \\
E_1 = iE_i(0)\sin(u) \,.
\eea
The diffraction efficiency after propagation from $z=0$ to $z=L$, $\eta=|E_1(L)|^2/|E_i(0)|^2$, can then be expressed as
\bea
\eta(z) = \sin^2 \left[ \frac{k}{2n_0}\int_0^z  n_1(z')dz' \right]  \,. \label{eq:eta}
\eea

For non-sinusoidal index variations along $x$ like the ones derived in the previous section, Eq. \eqref{eq:DN}, we have to decompose the index modulation into its Fourier series,
\bea
\Delta n(x,z) &=& \Delta n_0(z) + \sum_{m=1}^\infty n_m(z) \cos(mKx) \,, \\
n_m(z) &=& \frac{2}{\Lambda} \int_0^\Lambda \Delta n(x,z)\cos(mKx)dx  \,,  \label{eq:DNFourier}
\eea 
where $x=0$ was chosen as the location of the maxima of $\Delta T(x)$ in order to eliminate the sine functions from the series and keep only the cosines, and $\Delta n_0(z)=(1/\Lambda)\int_0^\Lambda \Delta n(x,z)dx$. Since the terms of the Fourier series decomposition correspond to harmonics of $\myb K$, and thus to the higher-order diffraction modes discussed above ($\myb k_m=\myb k_i-m\myb K, m\in \mathbb N$), the first-order term of the Fourier series decomposition is the only term contributing to the diffraction into the $m=1$ mode (Bragg diffraction regime) with $\rho\gg 1$ \cite{YarivBook}. %The index modulation $\Delta n(x,z)$ is obtained from the previous section, 

In other words: for $\rho\gg 1$, the diffraction formula Eq. \eqref{eq:eta} is valid for arbitrary (non-sinusoidal) index modulations along $x$, as long as one is careful to extract the first order mode $n_1(z)$ of the Fourier series decomposition of $\Delta n(x,z)$. The other assumptions from this diffraction model are:
\bi
\item a paraxial approximation for the diffracted wave, i.e., the diffracted beam must be near-collimated;
\item the angular divergence of the incident beam ($\delta \theta \approx 1/F_\#$, where $F_\#$ is the beam's F-number) must be smaller than the diffraction angle from the incident to the first-order diffracted mode $\theta_{0-1}\approx \lambda/\Lambda$, i.e.,
\bea
F_\# \gg \frac{\Lambda}{\lambda} \,; \label{eq:F}
\eea
\item a quasi-monochromatic diffracted wave, with a slowly-varying pulse shape compared to the laser frequency;
\item a grating that is uniform along $y$ and larger than the size of the diffracted beam.
\ei

These conditions are not merely for the validity of the theory, but also represent design constraints for achieving high diffraction efficiencies in experiments. Equation \eqref{eq:eta} is valid for arbitrary transverse intensity profiles of the diffracted beam, as long as the scales of the variations remain large compared to the laser wavelength. Practically, the diffracted beam should also have a wavelength larger than 300 nm in order to avoid absorption by ozone.

}

\begin{figure}[htbp]
\includegraphics{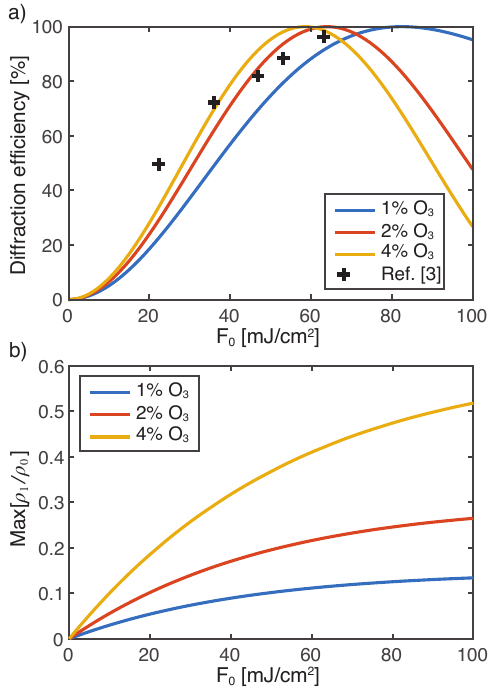}
\caption{a) Diffraction efficiency of a gas optics for conditions similar to Ref. \cite{MichineCP20}, and b) Maximum amplitude of the density modulation by the entrance of the gas optics ($z=1$ mm in Fig. \ref{fig:Inzt}), vs. initial average fluence of the UV laser and for an initial ozone fraction of 1\%, 2\% and 4\%.}
\label{fig:eta}
\end{figure}

Calculations of the diffraction efficiency for conditions similar to Ref. \cite{MichineCP20} (diffracted beam with $\lambda$ = 527 nm) are shown in Fig. \ref{fig:eta}a. Reasonable agreement is found with the experimental measurements, considering the approximations from our model and the experimental unknowns (the exact ozone fraction is not specified in Ref. \cite{MichineCP20}, apart from mentioning it was in the 1--10\% range). Figure \ref{fig:eta}b shows the maximum amplitude, defined here as half of peak-to-valley, vs. the average initial UV fluence. We have performed simulations of the 1D nonlinear Euler equations, Eqs. \eqref{eq:Euler1}--\eqref{eq:Euler2}, and verified that the expression connecting the index and temperature modulations derived in our linear analysis in Eq. \eqref{eq:DN}, when considering only the first-order mode of the Fourier series of the density $\rho_1$, remains valid until approximately $\rho_1/\rho_0\approx$ 40--50\%. Our choice of 4\% as the maximum ozone fraction in Fig. \ref{fig:eta} corresponds to the maximum ozone fraction for which our model assumptions remain valid.

The diffraction efficiency for a fixed and ``low'' UV fluence (before the curves reach $\eta=1$) increases rapidly with ozone fraction up to approximately 1\%, beyond which the efficiency becomes less sensitive to the ozone fraction. This is because the near-complete absorption of the UV light over the 1 cm of ozone occurs for an ozone fraction around 1\% for these fluences, where absorption saturation by ozone depletion is clearly visible but not sufficient for the laser to ``burn through'' the 1 cm of gas (cf. Fig. \ref{fig:Inzt}). Once the laser is fully absorbed, the effective diffraction length, which is the length over which most of the gas heating occurs (e.g., approximately 3--4 mm for 2\% ozone per Fig. \ref{fig:DTxz}a), keeps shrinking; the drop in diffraction length almost exactly compensates the increase in the index modulation from increased gas heating---cf. Eq.  \eqref{eq:eta}.

The increase in efficiency vs. fluence for a fixed ozone fraction can roll over after reaching $\eta=1$: this is the well-known oscillatory behavior of diffraction efficiency for volumetric transmission gratings, when the grating thickness exceeds the optimum value leading to $\eta=1$ (Eq. \eqref{eq:eta} for a constant $n_1$ simply gives $\eta(z)=\sin^2(k_i n_1 z/2n_0)$, cf. \cite{YarivBook,YehBook}).

{
To test the validity of our diffraction model, we have performed 3D wave propagation simulations using a paraxial wave solver (i.e., Fresnel propagation integral),
\bea
(2ik\dz + \nabla_\perp^2)E(\myb r) = -2k^2E(\myb r)\frac{\Delta n(\myb r)}{n_0} \,,
\eea
where $\Delta n(\myb r)$ is the 3D index modulation calculated in the previous section, Eq. \eqref{eq:DN}, with the parameters from Fig. \ref{fig:dNxz} (also red curve in Fig. \ref{fig:eta}a, for 2\% ozone fraction and $F_0$ = 60 mJ/cm$^2$). The propagation is solved in Fourier space along $(x,y)$, performing operator splitting between the diffraction and refraction steps along $z$. We calculate the propagation of a Gaussian laser beam with waist $w_0$ = 400 $\mu$m located at $z=0$, and used $\Lambda=60\lambda=31.6$ $\mu$m for the grating wavelength. The beam is incident at the Bragg angle onto the grating, $\theta_B\approx K/2k=\lambda/2\Lambda\approx$ 0.5\degree from the $z$-axis in the $(x,z)$ plane. The diffracted beam (mode 1) leaves at an angle of $-0.5$\degreens. Since the F-number of a Gaussian beam is given by $F_\#=\pi w_0/2\lambda \approx 1.2\times 10^3$ in our case, the condition from Eq. \eqref{eq:F} (with $\Lambda/\lambda=60$ here) is largely satisfied.

Figure \ref{fig:3DR}a shows the intensity of the diffracted beam in the ($x,z$) plane at $y=0$; the oxygen/ozone mixed gas is present between the two dashed green lines on the figure. Figure \ref{fig:3DR}b shows the power in the zero and first-order modes, from the simulations (after applying masks in Fourier space to isolate the diffraction modes) and from Eq. \eqref{eq:eta}. The difference between the simulations and theory is barely distinguishable on the plot. The modulated pattern along $x$ in Fig. \ref{fig:3DR}a comes from the interference between the zero and first-order modes; the contrast of the fringes is maximum near $z\approx$ 3 mm, which is where the two modes have approximately equal powers according to Fig. \ref{fig:3DR}b.

\begin{figure}[htbp]
\includegraphics{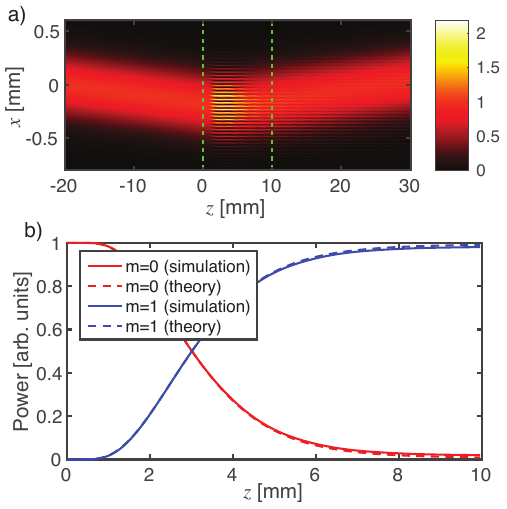}
\caption{3D paraxial wave simulation of the propagation of a Gaussian laser beam incident at the Bragg angle on the index modulation from Fig. \ref{fig:dNxz} with $\Lambda/\lambda=60$, i.e., $\rho$ = 4.7---diffraction in the Bragg regime, cf. Eq. \eqref{eq:rhodiff}: a) electric field intensity, showing the interference between the incident and first order diffracted modes (similar to the illustration from Fig. \ref{fig:setup}c; the oxygen/ozone gas is located between the two dashed green lines); b) power in the incident and first order modes vs. $z$, from the 3D simulation and from the coupled wave diffraction theory formula, Eq. \eqref{eq:eta}.}
\label{fig:3DR}
\end{figure}

The power spectrum of the electric field along the Fourier coordinates $k_x$, $k_y$ is shown in Fig. \ref{fig:3DF}a at $z=0$ and $z$ = 1 cm (the corresponding angles, in the small angle limit, are given by $\theta_x\approx k_x/k$, with $\theta_B\approx K/2k$), showing the spot moving from $k_x=K/2$ (mode 0, at $\theta_x=\theta_B=-$0.5\degreens) to $-K/2$ (mode 1, at $+0.5$\degreens). The line-out along $k_y=0$, shown on a logarithmic scale, reveals that almost all the incident power (in $m=0$) has been transferred into mode $m=1$, with $\approx$ 1\% of the incident power left in $m=0$. Mode $m=-1$ contains $\approx$ 0.4\% of the incident power; the power in $m=-1$ is the reason for the slight difference between theory and simulations in Fig. \ref{fig:3DR}b, since only $m=1$ is accounted for in the theory. This is consistent with the choice of $\Lambda/\lambda=60$, resulting in diffraction in the Bragg regime with $\rho=4.7$ per Eq. \eqref{eq:rhodiff}.

On the other hand, for $\rho\le 1$, diffraction into higher order modes can become significant. This is illustrated in Fig. \ref{fig:3DF}b, showing results from a simulation identical to \ref{fig:3DF}a except for a larger grating period, $\Lambda=200\lambda=105\, \mu$m, leading to $\rho=0.4$.

This illustrates that designing a gas grating like the one described here to use as a dielectric mirror (diffraction into the first order mode only) requires a careful choice of the grating wavelength $\Lambda$: on the one hand, shortening $\Lambda$ reduces the acoustic period $\tau_s=\Lambda/c_s$, which limits the maximum diffracted pulse duration that can be used while the density modulation is near its maximum amplitude at $t\approx \tau_s/2$. But on the other hand, lengthening $\Lambda$ can eventually lead to diffraction into higher order modes, as shown in Fig. \ref{fig:3DF}b.

%This condition might require some design compromise against the diffracted beam pulse duration: indeed, reducing $\Lambda$ leads to a shorter acoustic period $\tau_s=\Lambda/c_s$, which decreases the duration for which the index modulation is maximum near $t=\tau_s/2$ (cf. Fig. \ref{fig:wave}c). 

%higher order modes contain les than 0.1\% of the incident power, except for mode $m=-1$, which contains approximately 0.4\% of the incident power ()

%higher order diffraction modes are present after propagation through 1 cm, but contain a negligible amount of energy compared to mode 1. The integrated energy in the higher-order modes is at the origin of the very small difference between the theory curve and the simulation results in Fig. \ref{fig:3DR}b. The small contribution of the higher-order diffraction modes is consistent with our choice of grating wavelength $\Lambda=60\lambda$, leading to $\rho=4.7$ per Eq. \eqref{eq:rhodiff}, with $n_1\approx 6\times 10^{-5}$, i.e., diffraction in the Bragg regime. We have verified using similar simulations that the contribution of the higher-order modes becomes significant when the grating wavelength is increased up to $\rho\le 1$.

\begin{figure}[htbp]
\includegraphics{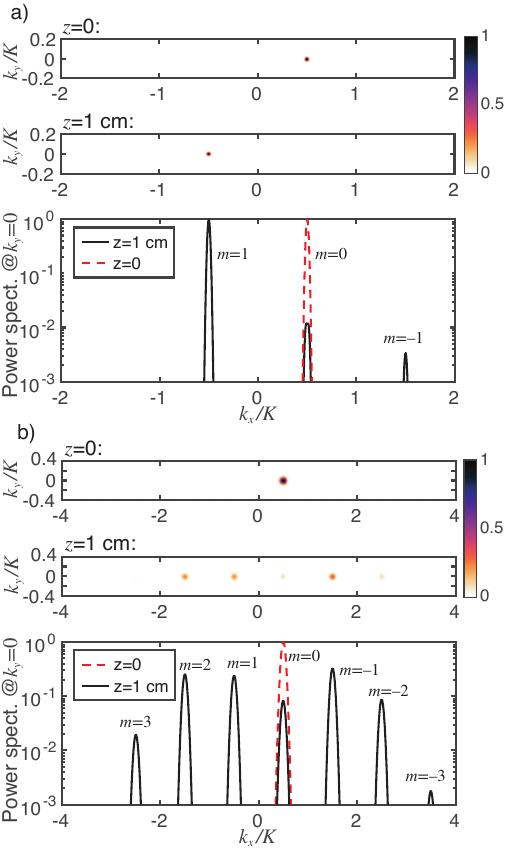}
\caption{Fourier analysis of 3D simulations, showing the transverse ($k_x,k_y$) Fourier profile of the total electric field at $z$= 0 and 1 cm, as well as lineouts (on a logarithmic scale) vs. $k_x$ at $k_y=0$: a) $\Lambda=60\lambda$, i.e., $\rho=4.7$ (same as Fig. \ref{fig:3DR}): diffraction in the Bragg regime, with near-complete transfer of power from the incident mode $m=0$ into $m=1$; b) $\Lambda=200\lambda$, i.e., $\rho=0.4$: the diffraction starts to enter the Raman-Nath regime, with significant energy diffracted into high-order modes ($m=-2$ to $+3$).}
\label{fig:3DF}
\end{figure}

}

\section{Discussion and future direction\label{sec:discuss}}%%%%%%%%%%%%%%%%%%%%%%%%

Table \ref{tab:reactions} and Fig. \ref{fig:DTxz}c) show that for the conditions of these experiments when the fraction of ozone is small, the most effective reaction besides the photodissociation is reaction 1 where O($^1$D) reacts with the oxygen buffer gas. The electronic energy of O($^1$D) mostly goes into electronic energy of O$_2$ ($^1\Sigma_g$ or $^1\Delta_g$), whose quenching rate is too slow to matter over the typical 10 ns duration of the UV imprint beam \cite{BurkholderNASA19}.

It should be possible to optimize the gas heating by choosing a different buffer gas. For example, we suggest using CO$_2$, or at least a mixture of O$_2$ and CO$_2$. This is motivated by the reaction of O($^1$D) with CO$_2$,
\bea
\text{O}(^1\text{D})+\text{CO}_2\rightarrow \text{O}+\text{CO}_2 \,,
\eea
i.e., the quenching of O($^1$D) back to its ground state via collisions with CO$_2$. The reaction enthalpy is therefore $-$1.97 eV, i.e., the full electronic energy of O($^1$D), of which 1.23 eV are available to heat the surrounding gas assuming energy equipartition (vs. 0.29 eV and 0.84 eV for reactions 1a and 1b), and the reaction rate, $k_6=1.1\times 10^{-10}$ cm$^3$/s \cite{BurkholderNASA19}, is also 3.4 and 13.8 times faster than for reactions 1a and 1b.

\begin{figure}[htbp]
\includegraphics{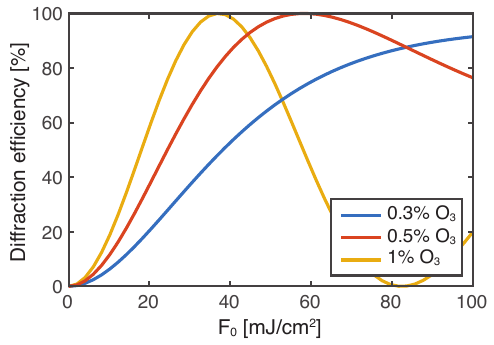}
\caption{Diffraction efficiency for a 1 cm gas optics (same as Fig. \ref{fig:eta}), but with CO$_2$ instead of O$_2$ used as the buffer gas.}
\label{fig:etaCO2}
\end{figure}

We estimate a significant increase in diffraction efficiency when replacing oxygen by carbon dioxide. Figure \ref{fig:etaCO2} shows the diffraction efficiency for the same parameters as Fig. \ref{fig:eta} (1 cm propagation length, room temperature) but with O$_2$ replaced by CO$_2$ as the buffer gas. Similar diffraction efficiencies are obtained for CO$_2$ compared to O$_2$ for only about a tenth of the ozone fraction---or, depending on the particular application and constraints, one can in principle reach high diffraction efficiency for shorter propagation distances, or a lower UV fluence.

Other reactions can potentially be utilized to increase the gas heating even further, like adding hydrogen to the buffer gas due to its fast reaction rate with O($^1$D) and the high exothermicity of the reaction.

Applications of gas optics for ICF or IFE experiments will require operations in vacuum environment, which might be the primary challenge facing the deployment of gas optics technology for these applications. Gas jets could in principle be utilized to generate gas densities close to those at atmospheric pressure; however, the high pressurization of ozone required in the process can lead to its rapid self-dissociation. Lowering the requirements on the fraction of ozone, like CO$_2$ is estimated to allow, should help find a workable parameter space for high-power lasers applications in vacuum environments.

{Finally, our model can easily be extended to the design of other volumetric holographic optical elements. In particular, a volumetric holographic lens, similar to the plasma lenses described in Refs. \cite{EdwardsPRL22,LehmannPRE19}, should be a relatively straightforward next step, requiring the extension of our 1D fluid model to both transverse dimensions ($x,y$).
}

\section{Conclusion}

In conclusion, we have derived a comprehensive description of the physics and chemistry of acousto-optics in gas initiated by the absorption of a modulated UV ``imprint'' beam and resulting gas heating. Such gas-based optical elements have the potential to transform the design and applications of high power lasers, including inertial fusion energy. {Compared to plasma gratings such as the ones described in our previous work \cite{EdwardsOptica23}, gas optics have lower index modulation amplitudes but significantly longer lifetimes (tens of ns vs. tens of ps for plasma gratings), making them better suited for long-pulse (ns) laser applications. } Our model gives good agreement with previous experiments and will be used to design future ones for high-power laser applications,
 { including for the design of other holographic elements relying on the same process, such as diffraction lenses.} Other configurations with different absorbing molecules or buffer gas are expected to further improve the performances of these novel optical elements.

\begin{acknowledgments}
We are grateful to Y. Michine, H. Yoneda, R. More, J. Wurtele, P. Cameron-Smith and L. Divol for fruitful discussions. This work was performed under the auspices of the U.S. Department of Energy by Lawrence Livermore National Laboratory under Contract DE-AC52-07NA27344, and was funded by the Laboratory Research and Development Program at LLNL under Project Tracking Code No. 24-ERD-001. This document was prepared as an account of work sponsored by an agency of the United States government. Neither the United States government nor Lawrence Livermore National Security, LLC, nor any of their employees makes any warranty, expressed or implied, or assumes any legal liability or responsibility for the accuracy, completeness, or usefulness of any information, apparatus, product, or process disclosed, or represents that its use would not infringe privately owned rights. Reference herein to any specific commercial product, process, or service by trade name, trademark, manufacturer, or otherwise does not necessarily constitute or imply its endorsement, recommendation, or favoring by the United States government or Lawrence Livermore National Security, LLC. The views and opinions of authors expressed herein do not necessarily state or reflect those of the United States government or Lawrence Livermore National Security, LLC, and shall not be used for advertising or product endorsement purposes.
\end{acknowledgments}

% to do here:
% - clean up DT(x)=DTmax*(1+cos(kx)) or DTmax/2*cos(kx) in Fig. 4 and text
% - figure out if sound speed goes up as sqrt(1+DTmax/T0) (Albertine); estimate error resulting from cs(z)

%To calculate the diffraction efficiency $\eta(z)$, we need to keep in mind that: i) the modulation amplitude of the refractive index is non-uniform along $z$; and ii) the transverse modulation profile at a fixed $z$ is generally non-sinusoidal, as mentioned in Sect. \ref{sec:DT} (cf. Fig. \ref{fig:DTxz}b).


\end{document}